\newcommand{\PRL}[3]{Phys.\ Rev.\ Lett.\ {\bf #1},\ #2 (#3)}

\newcommand{\NAT}[3]{Nature\ {\bf #1},\ #2 (#3)}

\newcommand{\PRA}[3]{Phys.\ Rev.\ A\ {\bf #1},\ #2 (#3)}

\newcommand{\JPA}[3]{J.\ Phys.\ A:\ Math.\ Gen.\ {\bf #1},\ #2 (#3)}

\newcommand{\JETP}[3]{JETP.\ Lett.\ {\bf #1},\ #2 (#3)}

\documentclass[aps,showkeys,showpacs,amsmath,amssymb]{revtex4}
\usepackage{longtable}
\begin{document}
\title{Quantum teleportation and state sharing using a genuinely entangled six qubit state}
\author{Sayan Choudhury}
\email{sayan.cho@gmail.com} 
\affiliation{Indian Institute of Science Education and Research (IISER) Kolkata, Salt Lake, Kolkata - 700106, India}
\author{Sreraman Muralidharan}
\email{sreraman_m@yahoo.co.in}
\affiliation{Loyola College, Nungambakkam, Chennai - 600034, India}
\author{Prasanta K. Panigrahi}
\email{prasanta@prl.res.in}
\affiliation{Indian Institute of Science Education and Research (IISER) Kolkata, Salt Lake, Kolkata - 700106, India}
\affiliation{Physical Research
Laboratory, Navrangpura, Ahmedabad - 380 009, India}

\begin{abstract}
The usefulness of the genuinely entangled six qubit state that was recently introduced by Borras \it et al. \normalfont
is investigated for the quantum teleportation of an arbitrary three qubit state and for quantum state sharing (QSTS) of an arbitrary two qubit state.
For QSTS, we explicitly devise two protocols and construct sixteen orthogonal measurement basis which can lock an arbitrary two qubit
information between two parties. 
\end{abstract}
\maketitle
\section{introduction}
Entanglement is the most striking and counter intuitive feature of quantum mechanics that has found
many practical applications in the field of cryptography and communication technology \cite{Nielson}. 
Entangled states like the Bell, GHZ and their generalizations play a significant role in
the accomplishment of various quantum tasks like teleportation \cite{Bennett}, secret sharing \cite{Gotit} and dense coding \cite{Wiesner}. It
is well characterized only upto four qubits \cite{Veres}. Intriguingly, not all entangled states are useful in carrying out the desired operations.
\\

In the case of three qubits, entanglement can be characterized into two in equivalent ways \cite{Cirac} : the GHZ and W state categories. 
While the GHZ states are suitable for carrying out various
quantum tasks, the normal W states \cite{Gorbo} are not. As is evident, the nature of the multipartite entanglement is crucial in determining the efficacy of the entangled state under consideration for quantum communication. The GHZ states have long range order characteristically different from the W state, for which such order is absent although it has greater local connectivity. \\

Bennett \it et al. \normalfont \cite{Bennett}, introduced the first scheme for the teleportation of an arbitrary single qubit state $|\psi_a\rangle = \alpha|0\rangle+\beta|1\rangle$, where
$\alpha, \beta \in C$ and $|\alpha|^2+|\beta|^2 = 1$ using an EPR pair as an entangled resource. 
This has been experimentally achieved both in laboratory, as well as in realistic conditions \cite{tele1, Bsm}.
Recently, attention has turned towards the investigation of the efficacy of a number of multipartite entangled channels for the
teleportation of an arbitrary two qubit state given by \cite{Rigolin1, Yeo, Sreraman, Sreraman2},
\begin{equation}
|\psi_b\rangle =  \alpha|00\rangle + \beta|10\rangle + \gamma|01\rangle + \delta|11\rangle
\end{equation}
where $\alpha, \beta, \gamma, \delta \in C$ and $|\alpha|^2+|\beta|^2+|\gamma|^2+|\delta|^2 = 1$.
Further, it was shown by two of the present authors
that it is possible to teleport an arbitrary $N$ qubit state of the type
\begin{equation}
|\psi_N\rangle = \sum_{i_1,i_2...i_n = 0}^{1} \alpha_{i_{1}i_{2}...i_{n}}|i_{1}i_{2}...i_{n} \rangle,  
\end{equation}
where $\alpha_{i_{1}i_{2}...i_{n}} \in C$ and $\Sigma|{\alpha_{i_{1}i_{2}...i_{n}}}|^2 = 1$,
using a $2N$ qubit state of the form \cite{Sreraman3}:
\begin{equation}
|\zeta_{2N}\rangle = \sum_{i_1,i_2...i_n = 0\\ i_1=i_2...=i_n \neq 1}^{1} R(i_{1}i_{2}...i_{n}) |i_{1}i_{2}...i_{n}\rangle - |1...1\rangle.
\end{equation}
Here $R$ refers to the unitary "Reflection operator" performing the transformation $|i_1,i_2...i_n\rangle \rightarrow |i_n...1_2,i_1\rangle$. \\

Quantum state sharing (QSTS) \cite{Qsts} is the use of teleportation for the secret sharing of quantum information
among various parties such that the reciever can obtain the required
information, only if all the members involved in the protocol cooperate. Hillery \it et al. \normalfont \cite{Hillery},
proposed the first scheme for the QSTS of a single qubit state $|\psi_a\rangle$ using a tri-partite GHZ
state. Later the usefulness of an asymmetic W state \cite{Zheng} was demonstrated for the same purpose and
experimentally realized in ion trap systems. \\

QSTS of a two qubit state $|\psi_b\rangle$ was
initially realized using four Bell pairs \cite{Deng}. Recently, two of the present
authors proposed that QSTS of $|\psi_b\rangle$ can be realized using the highly entangled five partite states
which are not decomposable into Bell pairs of the type \cite{Sreraman}:
\begin{equation}
|\psi_{5}\rangle=\frac{1}{2}(|\Omega_1\rangle|\phi_{-}\rangle+|\Omega_2\rangle|\psi_{-}\rangle
+|\Omega_3\rangle|\phi_{+}\rangle+|\Omega_4\rangle|\psi_{+}\rangle).
\end{equation}
where $|\Omega_i\rangle$'s form a tri-partite orthogonal basis. The same has been achieved by using the cluster state \cite{Sreraman4} 
\begin{equation}
|C_N\rangle = \frac{1}{2^{N/2}} \otimes_{a=1}^{N} (|0\rangle_a \sigma_z ^{a+1} + |1\rangle_a), 
\end{equation}
with $\sigma_z^{N+1} = 1$. In the experimental realization of multi-partite entangled states \cite{Lu1} one often starts with multiple copies of the Bell states which are subsequently further entangled. In an analogous manner, theoretical search for multi-partite entangled states often takes recourse to assembling of the desired state from constituents of Bell, GHZ states etc. It is worth observing that the construction of higher dimensional states relies on computational optimisation schemes \cite{Brown} and may not be familiar from physical considerations. Therefore, the efficacy of these states needs to be checked with several quantum protocols. \\ Borras \it et al.\normalfont \ \cite{Borras}, introduced a genuinely entangled six qubit state which is not decomposable
into pairs of Bell states.  It is given by, 
\begin{eqnarray}
|\psi_{6} \rangle  = \frac{1}{4} [|000\rangle(|0\rangle|\phi_{+}\rangle + |1\rangle|\psi_{+}\rangle) 
+ |001\rangle(|0\rangle|\psi_{-}\rangle - |1\rangle|\phi_{-}\rangle)
+ |010\rangle(|0\rangle|\psi_{+}\rangle - |1\rangle|\phi_{+}\rangle)
+ |011\rangle(|0\rangle|\phi_{-}\rangle + |1\rangle|\psi_{-}\rangle) \nonumber \\
+ |100\rangle(-|0\rangle|\psi_{-}\rangle - |1\rangle|\phi_{-}\rangle)
+|101\rangle(-|0\rangle|\phi_{+}\rangle + |1\rangle|\psi_{+}\rangle)   
+ |110\rangle(|0\rangle|\phi_{-}\rangle - |1\rangle|\psi_{-}\rangle)
+ |111\rangle(|0\rangle|\psi_{+}\rangle + |1\rangle|\phi_{+}\rangle)]. 
\end{eqnarray}
This state exhibits genuine entanglement according to many measures. The reduced single, two and three qubit
density matrices of this state are all completely mixed. Further it has been pointed out that no other pure state of six
qubits has been found that evolves to a mixed state with a higher amount of entanglement \cite{Borras2}. The state is robust against
decoherence and its entanglement still prevails after particle loss. This state also satisfies the monogamy inequality given by \cite{mono},
\begin{equation}
\sum_{i=2}^{n} C^2_{A_1 A_i} \leq C^2_{A_1|A_2...A_n},
\end{equation}
where $C_{A|B}$, represents the concurrence between the subsystems $A$ and $B$. 
 Hence, this state turns out to be an important resource for quantum communication protocols.
 Here, we show that this state can be used for teleportation of an arbitrary three qubit state and 
 for the QSTS of an arbitrary two qubit state in two distinct ways.  
\section{Teleportation}
Let Alice and Bob have the first three and the last three qubits in $|\psi_6\rangle$ respectively.
Alice has an arbitrary three qubit state given by,
\begin{equation}
|\psi_3\rangle = \sum_{i_1,i_2,i_3 = 0}^{1} \alpha_{i_{1}i_{2}i_{3}}|i_{1}i_{2}i_{3} \rangle,  
\end{equation}
where $\alpha_{i_{1}i_{2}i_{3}} \in C$ and $\Sigma|{\alpha_{i_{1}i_{2}i_{3}}}|^2 = 1$ which
she wants to teleport to Bob. Now, Alice can combine the the above state her part of the entangled state and perform a
six qubit measurement on her qubits and convey the outcome of her measurement to Bob via six cbits. For instance if Alice measures,
\begin{equation}
\sum_{i_1,i_2...i_3 = 0} ^{1} |i_{1}i_{2}i_{3}\rangle |i_{1}i_{2}i_{3}\rangle,
\end{equation}
then Bob's system collapses to $\sum \alpha_{i_1,i_2,i_3}|\zeta_{i_1 i_2 1_3}\rangle$, where $|\zeta_{i_1 i_2 1_3}\rangle$'s
are given by, 

\begin{eqnarray}
|\zeta_{000}\rangle &=& \frac{1}{\sqrt 2}(|0\rangle|\phi_{+}\rangle + |1\rangle|\psi_{+}\rangle), \\
|\zeta_{001}\rangle &=& \frac{1}{\sqrt 2}(|0\rangle|\psi_{-}\rangle - |1\rangle|\phi_{-}\rangle), \\ 
|\zeta_{010}\rangle &=& \frac{1}{\sqrt 2}(|0\rangle|\psi_{+}\rangle - |1\rangle|\phi_{+}\rangle), \\ 
|\zeta_{011}\rangle &=& \frac{1}{\sqrt 2}(|0\rangle|\phi_{-}\rangle + |1\rangle|\psi_{-}\rangle), \\
|\zeta_{100}\rangle &=&\frac{1}{\sqrt 2}(-|0\rangle|\psi_{-}\rangle - |1\rangle|\phi_{-}\rangle), \\
|\zeta_{101}\rangle &=&\frac{1}{\sqrt 2}(-|0\rangle|\phi_{+}\rangle + |1\rangle|\psi_{+}\rangle), \\
|\zeta_{110}\rangle &=& \frac{1}{\sqrt 2}(|0\rangle|\phi_{-}\rangle - |1\rangle|\psi_{-}\rangle), \\
|\zeta_{111}\rangle &=& \frac{1}{\sqrt 2}(|0\rangle|\psi_{+}\rangle + |1\rangle|\phi_{+}\rangle).
\end{eqnarray}
Now, Bob can perform an appropriate unitary operation on his qubits and 
obtain $|\psi_6\rangle$. We shall investigate the usefulness of this state
for QSTS of an arbitrary two qubit state in the forthcoming sections. 
\section{QSTS of an arbitrary two qubit state}
\subsection{Protocol 1}
We let Alice possess particles 1,2, Bob possess particles 3,4, and Charlie
possess particles 5 and 6 in $|\psi_6\rangle$ respectively. Alice combines
the state $|\psi_{b}\rangle$ with $|\psi_{6}\rangle$
and performs a four - particle measurement and conveys the outcome
of her measurement to Charlie by four cbits of information. The outcome
of the measurement made by Alice and the entangled state obtained
by Bob and Charlie are shown in the following table :-
\begin{table}[h]
\caption{\label{tab9} The outcome of the measurement performed by Alice and the state obtained by Bob
and Charlie(The coefficient $\frac{1}{2}$ is removed for convenience)}
\begin{tabular}{|c|c|}
\hline {\bf Outcome of the Measurement } & {\bf State obtained}\\
\hline
$|0000\rangle+|1001\rangle\pm|0111\rangle\pm|1110\rangle$&$\alpha|\eta_{1}\rangle+\mu|\eta_{2}\rangle\pm\gamma|\eta_{3}\rangle\pm\beta|\eta_{4}\rangle$\\
$|0000\rangle-|1001\rangle\pm|0111\rangle\mp|1110\rangle$&$\alpha|\eta_{1}\rangle-\mu|\eta_{2}\rangle\pm\gamma|\eta_{3}\rangle\mp\beta|\eta_{4}\rangle$\\
$|0001\rangle+|1000\rangle\pm|0110\rangle\pm|1111\rangle$&$\alpha|\eta_{2}\rangle+\mu|\eta_{1}\rangle\pm\gamma|\eta_{4}\rangle\pm\beta|\eta_{3}\rangle$\\
$|0001\rangle-|1000\rangle\pm|0110\rangle\mp|1111\rangle$&$\alpha|\eta_{2}\rangle-\mu|\eta_{1}\rangle\pm\gamma|\eta_{4}\rangle\mp\beta|\eta_{3}\rangle$\\
$|0011\rangle+|1010\rangle\pm|0100\rangle\pm|1101\rangle$&$\alpha|\eta_{3}\rangle+\mu|\eta_{4}\rangle\pm\gamma|\eta_{1}\rangle\pm\beta|\eta_{2}\rangle$\\
$|0011\rangle-|1010\rangle\pm|0100\rangle\mp|1101\rangle$&$\alpha|\eta_{3}\rangle-\mu|\eta_{4}\rangle\pm\gamma|\eta_{1}\rangle\mp\beta|\eta_{2}\rangle$\\
$|0010\rangle+|1011\rangle\pm|0101\rangle\pm|1100\rangle$&$\alpha|\eta_{4}\rangle+\mu|\eta_{3}\rangle\pm\gamma|\eta_{2}\rangle\pm\beta|\eta_{1}\rangle$\\
$|0010\rangle-|1011\rangle\pm|0101\rangle\mp|1100\rangle$&$\alpha|\eta_{4}\rangle-\mu|\eta_{3}\rangle\pm\gamma|\eta_{2}\rangle\mp\beta|\eta_{1}\rangle$\\
\hline
\end{tabular}
\end{table} \\
Here the $|\eta_i\rangle's$ are given by :
\begin{eqnarray}
|\eta_{1}\rangle&=&\frac{1}{2}(|\phi_{-}\rangle|00\rangle + |\phi_{+}\rangle|11\rangle + |\psi_{+}\rangle|01\rangle + |\psi_{-}\rangle|10\rangle), \\
|\eta_{2}\rangle&=&\frac{1}{2}(-|\psi_{-}\rangle|00\rangle -|\psi_{+}\rangle|11\rangle + |\phi_{+}\rangle|01\rangle + |\phi_{-}\rangle|10\rangle), \\
|\eta_{3}\rangle&=&\frac{1}{2}(|\phi_{+}\rangle|00\rangle - |\phi_{-}\rangle|11\rangle - |\psi_{-}\rangle|01\rangle + |\psi_{+}\rangle|10\rangle),\\
|\eta_{4}\rangle&=&\frac{1}{2}(-|\psi_{+}\rangle|00\rangle +|\psi_{-}\rangle|11\rangle + |\phi_{+}\rangle|10\rangle - |\phi_{-}\rangle|01\rangle). 
\end{eqnarray} 
It could be noticed that each four partite measurement basis in table 1 could be further broken down into Bell and single partite measurements
. For instance the first measurement basis could be written as
\begin{equation}
(|\psi_+\rangle (|0\rangle+|1\rangle) + |\psi_{-}\rangle (|0\rangle-|1\rangle)) |0\rangle + (|\phi_{-}\rangle (|0\rangle-|1\rangle) +
|\phi_+\rangle (|0\rangle+|1\rangle)) |1\rangle,
\end{equation}
where $|\psi_+\rangle$, $|\psi_{-}\rangle$, $|\phi_+\rangle$ and $|\phi_{-}\rangle$ refers to the Bell states respectively.
Bob can perform a two qubit measurement on his qubits and communicate the outcome of his measurement to Charlie who then, performs an
appropriate unitary transformation to get the state $|\psi_b\rangle$. Suppose, the Bob-Charlie system collapses to $\alpha|\eta_{1}\rangle+\mu|\eta_{2}\rangle+\gamma|\eta_{3}\rangle+\beta|\eta_{4}\rangle$, then if Bob wants to perform a Bell measurement,  
the outcome of the measurement performed by Bob and the state received by Charlie is shown in table II :\\
\begin{table}[h]
\caption{\label{tab6} The outcome of the measurement performed by Bob and the state obtained Charlie.}
\begin{tabular}{|c|c|}
\hline {\bf Outcome of the Measurement } & {\bf State obtained }\\
$|\phi_+\rangle$&$\alpha|11\rangle-\gamma |00\rangle+\beta|10\rangle+\mu|01\rangle$\\
$|\phi_{-}\rangle$&$\alpha|00\rangle-\gamma|11\rangle-\beta|01\rangle+\mu|10\rangle$\\
$|\psi_+\rangle$&$\alpha|01\rangle+\gamma|10\rangle-\beta|00\rangle-\mu|11\rangle$\\
$|\psi_{-}\rangle$&$\alpha|10\rangle-\gamma|01\rangle+\beta|11\rangle-\mu|00\rangle$\\
\hline
\end{tabular}
\end{table}\\
Instead of choosing the above measurement basis, if Bob chooses to measure in the computational basis, then the outcome of the measurement performed by Bob and the state obtained by Charlie is shown in the table III :\\
\begin{table}[h]
\caption{\label{tab6} The outcome of the measurement performed by Bob and the state obtained Charlie.}
\begin{tabular}{|c|c|}
\hline {\bf Outcome of the Measurement } & {\bf State obtained }\\
$|00\rangle$&$\alpha|\phi_+\rangle+\gamma |\phi_{-}\rangle+\mu|\psi_+\rangle-\beta\psi_{-}\rangle$\\
$|01\rangle$&$\alpha|\psi_+\rangle-\gamma |\psi_{-}\rangle-\mu|\phi_{+}\rangle-\beta|\phi_{-}\rangle$\\
$|10\rangle$&$\alpha|\psi_{-}\rangle+\gamma|\psi_+\rangle+\mu|\phi_{-}\rangle-\beta|\phi_+\rangle$\\
$|11\rangle$&$-\alpha|\phi_{-}\rangle+\gamma|\phi_+\rangle+\mu|\psi_{-}\rangle+\beta|\psi_+\rangle$\\
\hline
\end{tabular}
\end{table}
\subsection {Protocol 2}
We can also demonstrate a different protocol for the QSTS of $|\psi_b\rangle$ using $|\psi_6\rangle$ as an entangled resource by redistributing
the particles among Alice, Bob and Charlie. In this protocol, we let Alice possess particles 1,2 and 3; Bob possess particle 4 and Charlie possess particles 5 and 6 in $|\psi_6\rangle$ respectively. Alice first, combines the state $|\psi_{b}\rangle$ with $|\psi_{6}\rangle$, performs a five particle measurement and conveys the outcome of her measurement to Charlie by three cbits of information. 
The measurement performed by Alice and the corresponding entangled states obtained by Bob and Charlie
are shown in the table IV :
\begin{table}[h]
\caption{\label{tab9} The outcome of the measurement performed by Alice and the state obtained by Bob
and Charlie(The coefficient $\frac{1}{2}$ is removed for convenience)}
\begin{tabular}{|c|c|}
\hline {\bf Outcome of the Measurement } & {\bf State obtained}\\
\hline
$|00000\rangle+|10001\rangle\pm|01011\rangle\pm|11010\rangle$&$\alpha|\zeta_{1}\rangle+\mu|\zeta_{2}\rangle\pm\gamma|\zeta_{4}\rangle\pm\beta|\zeta_{3}\rangle$\\
$|00000\rangle-|10001\rangle\pm|01011\rangle\mp|11010\rangle$&$\alpha|\zeta_{1}\rangle-\mu|\zeta_{2}\rangle\pm\gamma|\zeta_{4}\rangle\mp\beta|\zeta_{3}\rangle$\\
$|00010\rangle+|10011\rangle\pm|01101\rangle\pm|11100\rangle$&$\alpha|\zeta_{3}\rangle+\mu|\zeta_{4}\rangle\pm\gamma|\zeta_{6}\rangle\pm\beta|\zeta_{5}\rangle$\\
$|00010\rangle-|10011\rangle\pm|01101\rangle\mp|11100\rangle$&$\alpha|\zeta_{3}\rangle-\mu|\zeta_{4}\rangle\pm\gamma|\zeta_{6}\rangle\mp\beta|\zeta_{5}\rangle$\\
$|00110\rangle+|10111\rangle\pm|01001\rangle\pm|11000\rangle$&$\alpha|\zeta_{7}\rangle+\mu|\zeta_{8}\rangle\pm\gamma|\zeta_{2}\rangle\pm\beta|\zeta_{1}\rangle$\\
$|00110\rangle-|10111\rangle\pm|01001\rangle\mp|11000\rangle$&$\alpha|\zeta_{7}\rangle-\mu|\zeta_{8}\rangle\pm\gamma|\zeta_{2}\rangle\mp\beta|\zeta_{1}\rangle$\\
$|00100\rangle+|10101\rangle\pm|01010\rangle\pm|11011\rangle$&$\alpha|\zeta_{5}\rangle+\mu|\zeta_{6}\rangle\pm\gamma|\zeta_{3}\rangle\mp\beta|\zeta_{4}\rangle$\\
$|00100\rangle-|10101\rangle\pm|01010\rangle\mp|11011\rangle$&$\alpha|\zeta_{5}\rangle-\mu|\zeta_{6}\rangle\pm\gamma|\zeta_{3}\rangle\mp\beta|\zeta_{4}\rangle$\\
\hline
\end{tabular}
\end{table}

Here the $|\zeta_{i}\rangle$'s are given by ;
\begin{eqnarray}
|\zeta_{1}\rangle &=& \frac{1}{\sqrt 2}(|0\rangle|\phi_{+}\rangle + |1\rangle|\psi_{+}\rangle),\\
|\zeta_{2}\rangle &=& \frac{1}{\sqrt 2}(|0\rangle|\psi_{-}\rangle - |1\rangle|\phi_{-}\rangle),\\
|\zeta_{3}\rangle &=& \frac{1}{\sqrt 2}(|0\rangle|\psi_{+}\rangle - |1\rangle|\phi_{+}\rangle),\\
|\zeta_{4}\rangle &=& \frac{1}{\sqrt 2}(|0\rangle|\phi_{-}\rangle + |1\rangle|\psi_{-}\rangle),\\
|\zeta_{5}\rangle &=& \frac{1}{\sqrt 2}(-|0\rangle|\psi_{-}\rangle - |1\rangle|\phi_{-}\rangle),\\
|\zeta_{6}\rangle &=& \frac{1}{\sqrt 2}(-|0\rangle|\phi_{+}\rangle + |1\rangle|\psi_{+}\rangle),\\
|\zeta_{7}\rangle &=& \frac{1}{\sqrt 2}(|0\rangle|\phi_{-}\rangle - |1\rangle|\psi_{-}\rangle),\\
|\zeta_{8}\rangle &=& \frac{1}{\sqrt 2}(|0\rangle|\psi_{+}\rangle + |1\rangle|\phi_{+}\rangle).\\
\end{eqnarray}\\ 
Now, Bob can perform a measurement in the basis ${|0\rangle, |1\rangle}$ and convey the result of his measurement to Charlie after which, Charlie can
apply an appropriate unitary transformation on his qubits to get back $|\psi_b\rangle$. For instance if the combined
state of Bob and Charlie collapses to the third state given in table IV, then the outcome of the measurement performed
by Bob and the state obtained by Charlie is shown in table V :  
\begin{table}[h]
\caption{\label{tab7} The outcome of the measurement performed by Bob and the state obtained by Charlie}
\begin{tabular}{|c|c|}
\hline {\bf Outcome of the Measurement } & {\bf State obtained }\\
$|0\rangle$&$\alpha|\psi_{+}\rangle+\mu|\phi_{-}\rangle\mp\gamma|\phi_{+}\rangle\mp\beta|\psi_{-}\rangle$\\
$|1\rangle$&$-\alpha|\phi_{+}\rangle+\mu|\psi_{-}\rangle\pm\gamma|\psi_{+}\rangle\mp\beta|\phi_{-}\rangle$\\
\hline
\end{tabular}
\end{table}\\
Now Charlie can apply an appropriate unitary operator on his qubits and get back $|\psi_b\rangle$.
Subsequently, Alice can also choose a different basis set, in which case Bob can measure in the Hadamard basis. The result of the measurement performed by Alice and the combined state of Bob and Charlie is shown in the table VI :\\
\begin{table}[h!]
\caption{\label{tab9} The outcome of the measurement performed by Alice and the state obtained by Bob
and Charlie(The coefficient $\frac{1}{2}$ is removed for convenience)}
\begin{tabular}{|c|c|}
\hline {\bf Outcome of the Measurement } & {\bf State obtained}\\
\hline
$|00000\rangle+|10001\rangle\pm|01011\rangle\pm|11010\rangle$&$\alpha|\zeta_{1}\rangle+\mu|\zeta_{2}\rangle\pm\gamma|\zeta_{4}\rangle\pm\beta|\zeta_{3}\rangle$\\
$|00000\rangle-|10001\rangle\pm|01011\rangle\mp|11010\rangle$&$\alpha|\zeta_{1}\rangle-\mu|\zeta_{2}\rangle\pm\gamma|\zeta_{4}\rangle\mp\beta|\zeta_{3}\rangle$\\
$|00101\rangle+|10100\rangle\pm|01111\rangle\pm|11110\rangle$&$\alpha|\zeta_{6}\rangle+\mu|\zeta_{5}\rangle\pm\gamma|\zeta_{8}\rangle\pm\beta|\zeta_{7}\rangle$\\
$|00101\rangle-|10100\rangle\pm|01111\rangle\mp|11110\rangle$&$\alpha|\zeta_{6}\rangle-\mu|\zeta_{5}\rangle\pm\gamma|\zeta_{8}\rangle\mp\beta|\zeta_{7}\rangle$\\
$|00110\rangle+|10111\rangle\pm|01010\rangle\pm|11011\rangle$&$\alpha|\zeta_{7}\rangle+\mu|\zeta_{8}\rangle\pm\gamma|\zeta_{3}\rangle\pm\beta|\zeta_{4}\rangle$\\
$|00110\rangle-|10111\rangle\pm|01010\rangle\mp|11011\rangle$&$\alpha|\zeta_{7}\rangle-\mu|\zeta_{8}\rangle\pm\gamma|\zeta_{3}\rangle\mp\beta|\zeta_{4}\rangle$\\
$|00100\rangle+|10101\rangle\pm|01001\rangle\pm|11000\rangle$&$\alpha|\zeta_{5}\rangle+\mu|\zeta_{6}\rangle\pm\gamma|\zeta_{2}\rangle\pm\beta|\zeta_{1}\rangle$\\
$|00100\rangle-|10101\rangle\pm|01001\rangle\mp|11000\rangle$&$\alpha|\zeta_{5}\rangle-\mu|\zeta_{6}\rangle\pm\gamma|\zeta_{2}\rangle\mp\beta|\zeta_{1}\rangle$\\
\hline
\end{tabular}
\end{table} 

Bob can perform a measurement in the basis $\frac{1}{\sqrt 2}(|0\rangle \pm |1\rangle)$ and convey the result of his measurement to Charlie. 
Now, Charlie can apply an unitary transformation on his qubits to get the state $|\psi_b\rangle$. For instance had the Bob-Charlie system collapsed into the state $\alpha|\zeta_{7}\rangle+\mu|\zeta_{8}\rangle+\gamma|\zeta_{3}\rangle+\beta|\zeta_{4}\rangle$, then the outcome of the
measurements performed by Bob and the corresponding states obtained by Charlie are shown in table VII :
\begin{table}[h]
\caption{\label{tab7} The outcome of the measurement performed by Bob and the state obtained by Charlie}
\begin{tabular}{|c|c|}
\hline {\bf Outcome of the Measurement } & {\bf State obtained }\\
$\frac{1}{\sqrt 2}(|0\rangle + |1\rangle)$&$\alpha(|\phi_{-}\rangle-|\psi_{-}\rangle)+\mu(|\phi_{+}\rangle+|\psi_{+}\rangle)+\gamma(|\psi_{+}\rangle-|\phi_{+}\rangle) +\beta(|\psi_{-}\rangle+|\phi_{-}\rangle)$\\
$\frac{1}{\sqrt 2}(|0\rangle - |1\rangle)$&$\alpha(|\phi_{-}\rangle+|\psi_{-}\rangle)-\mu(|\phi_{+}\rangle-|\psi_{+}\rangle)+\gamma(|\psi_{+}\rangle+|\phi_{+}\rangle) -\beta(|\psi_{-}\rangle-|\phi_{-}\rangle)$\\
\hline
\end{tabular}
\end{table}

\section{Conclusion}
In this paper, we have demonstrated the usefulness of a recently introduced six qubit state for the teleportation
of an arbitrary three qubit state and for the quantum state sharing of an arbitrary
two qubit state in two distinct ways. Further, this state satisfies the conjecture made by two of the present authors \cite{Sreraman4}
that the number of distinct ways in which one can split an arbitrary n qubit state using a genuinely entangled N
qubit state as an entangled channel, among two parties in the case where they need not meet up is $(N-2n)$. The spectacular 
properties of this state makes our protocols robust against decoherence. In future, we wish to study these protocols through nosiy channels and
investigate the decoherence properties of this state.


\begin{thebibliography}{26}
\bibitem{Nielson} M. A. Nielsen and I. L. Chuang, {\it Quantum Computation and Quantum Information,} (Cambridge Univ. Press, 2002).
\bibitem{Bennett} C. H. Bennett, G. Brassard, C. Crepeau, R. Jozsa, A. Peres, and W. K. Wootters,
 \PRL{70}{1895}{1993}.
\bibitem{Gotit} D. Gottesman, \PRA{61}{042311}{2000}.
\bibitem{Wiesner} C. H. Bennett and S. J. Wiesner, \PRL{69}{2881}{1992}.
\bibitem{Veres} F. Verstraete, J. Dehaene, B. DeMoor, and H. Verschelde, \PRA{65}{052112}{2002}.
\bibitem{Cirac} W. Dur, G. Vidal, and J. I. Cirac, \PRA{62}{062314}{2000}.
\bibitem{Gorbo} V. N. Gorbachev and A. I. Trubilko, \JETP{91}{894}{2000}.
\bibitem{Bennett} C. H. Bennett, G. Brassard, C. Crepeau, R. Jozsa, A. Peres, and W. K. Wootters,
 \PRL{70}{1895}{1993}.
\bibitem{tele1} D. Bouwmeester, J. W. Pan, K. Mattle, M. Eibl, H. Weinfurter, and A. Zeilinger
\NAT{390}{575}{1997}.
\bibitem{Bsm} Q. Zhang, A. Goebel, C. Wagenknecht, Y. A. Chen, B. Zhao, T. Yang, A. Mair, J. Schmiedmayer and J. W. Pan,
\NAT{2}{678}{2006}.
\bibitem{Rigolin1} G. Rigolin, Phys. Rev. A 71 (2005) 032303.
\bibitem{Yeo} Y. Yeo and W. K. Chua, Phys. Rev. Lett. 96 (2006) 060502.
\bibitem{Sreraman} S. Muralidharan and P. K. Panigrahi, Phys. Rev. A 77 (2008) 032321.
\bibitem{Sreraman2} S. Muralidharan and P. K. Panigrahi, eprint quant-ph/0802.3484.
\bibitem{Sreraman3} S. Muralidharan, S. Karumanchi and P. K. Panigrahi, eprint quant-ph/0804.4206v2.
\bibitem{Qsts} A.M. Lance, T. Symul, W.P. Bowen, B.C. Sanders, and P.K. Lam, \PRL{92}{177903}{2004}.
\bibitem{Hillery} M. Hillery, V. Buzek, and A. Berthiaume, Phys. Rev. A 59 (1999) 1829.
\bibitem{Zheng} S. B. Zheng, \PRA {74}, {054303} {2006}.
\bibitem{Deng} F. G. Deng, X. H. Li, C. Y. Li, P. Zhou, and H. Y. Zhou, \PRA {72}, {044301} {2005}.
\bibitem{Sreraman4} S. Muralidharan and P. K. Panigrahi, eprint quant-ph/0802.0781v2.
\bibitem{Lu1} C. Y. Lu, X. Q. Zhou, O. Ghne, W. B. Gao, J. Zhang, Z. S. Yuan, A. Goebel, T. Yang, J. W. Pan
\NAT{3}{91}{2007}.
\bibitem{Brown} I. D. K. Brown, S. Stepney, A. Sudbery, and S. L. Braunstein,  \JPA {38} {1119} {2005}.
\bibitem{Borras} A. Borras, A. R. Plastino, J. Batle, C. Zander, M. Casas, and A. Plastino,  \JPA {40} {13407} {2007}.
\bibitem{Borras2} A. Borras, A.P. Majtey, A.R. Plastino, M. Casas, A. Plastino, eprint quant-ph/0806.0779v2.
\bibitem{mono} V. Coffman, J. Kundu and W.K. Wootters, \PRA{61} {052306} {2000}. 
\end{thebibliography}
\end{document}